\documentclass[notitlepage,showpacs,amsmath,amssymb,aps,prl,twocolumn]{revtex4-1}

\usepackage{amsmath,amssymb}
\usepackage{graphicx}
\usepackage{dcolumn}
\usepackage{bm}
\usepackage{color}
\usepackage{times}
\usepackage{wasysym}
\usepackage{times}
\usepackage{multirow}
\usepackage{verbatim}
\usepackage{subcaption}
\usepackage{caption}

\renewcommand{\vec}[1]{\mathbf{#1}}

\renewcommand{\eqref}[1]{(\ref{eq:#1})}

\newcommand{\figref}[1]{Fig.~\ref{fig:#1}}
\newcommand{\Figref}[1]{Figure~\ref{fig:#1}}
\newcommand{\figreftwo}[2]{Figs.~\ref{fig:#1} and~\ref{fig:#2}}

\newcommand{\beginsupplement}{%
        \setcounter{table}{0}
        \renewcommand{\thetable}{S\arabic{table}}%
        \setcounter{figure}{0}
        \renewcommand{\thefigure}{S\arabic{figure}}%
        \setcounter{equation}{0}
        \renewcommand{\theequation}{S\arabic{equation}}%
     }

\begin{document}

\title{Topology Optimized Multi-layered Meta-optics}

\author{Zin Lin$^1$}
\email{zinlin@g.harvard.edu}
\author{Benedikt Groever$^1$}
\author{Federico Capasso$^1$}
\author{Alejandro W. Rodriguez$^2$}
\author{Marko Lon\v{c}ar$^1$}
\affiliation{$^1$John A. Paulson School of Engineering and Applied Sciences, Harvard University, Cambridge, MA 02138}
\affiliation{$^2$Department of Electrical Engineering, Princeton University, Princeton, NJ, 08544}

\date{\today}

\begin{abstract}
  We propose a general topology optimization framework for metasurface
  inverse design that can automatically discover highly complex
  multi-layered meta-structures with increased functionalities. In
  particular, we present topology-optimized multi-layered geometries
  exhibiting angular phase control, including a single-piece
  nanophotonic metalens with angular aberration correction as well as
  an angle-convergent metalens that focuses light onto the same
  focal spot regardless of the angle of incidence.
\end{abstract}

\maketitle

Phase-gradient metasurfaces~\cite{yu2011light} have recently received
widespread attention due to their successful applications in important
technologies such as beam steering, imaging and
holography~\cite{aieta2015multiwavelength,arbabi2015dielectric,khorasaninejad2016metalenses}. Although
they offer many advantages in terms of size and scaling over
traditional refractive bulk optics, their capabilities are limited
with respect to spectral and angular
control~\cite{estakhri2016wave,arbabi2017fundamental}. Theoretical
analysis of ultra-thin metasurfaces suggests that to circumvent such
limitations, it might be necessary to employ exotic elements such as
active permittivities (e.g. optical gain), bi-anisoptropy, magnetic
materials, or even nonlocal response~\cite{estakhri2016wave}. Although
materials with such properties might be found in the radio-frequency
(RF) regime, they are not readily available at optical
frequencies. Alternatively, device functionalities may be enhanced
by increasingly complex geometric design. For instance,
multi-functional devices have been demonstrated by cascading a few
layers of metasurfaces, each of which comprises typical dielectric
materials~\cite{arbabi2014planar,arbabi2016miniature}. So far, most of
these multi-layered meta-structures (MMS) fall into a category of
structures where each layer is sufficiently far apart from the other
and can be considered independently.

\begin{figure}[t]
\centering
\includegraphics[width=1.0\columnwidth]{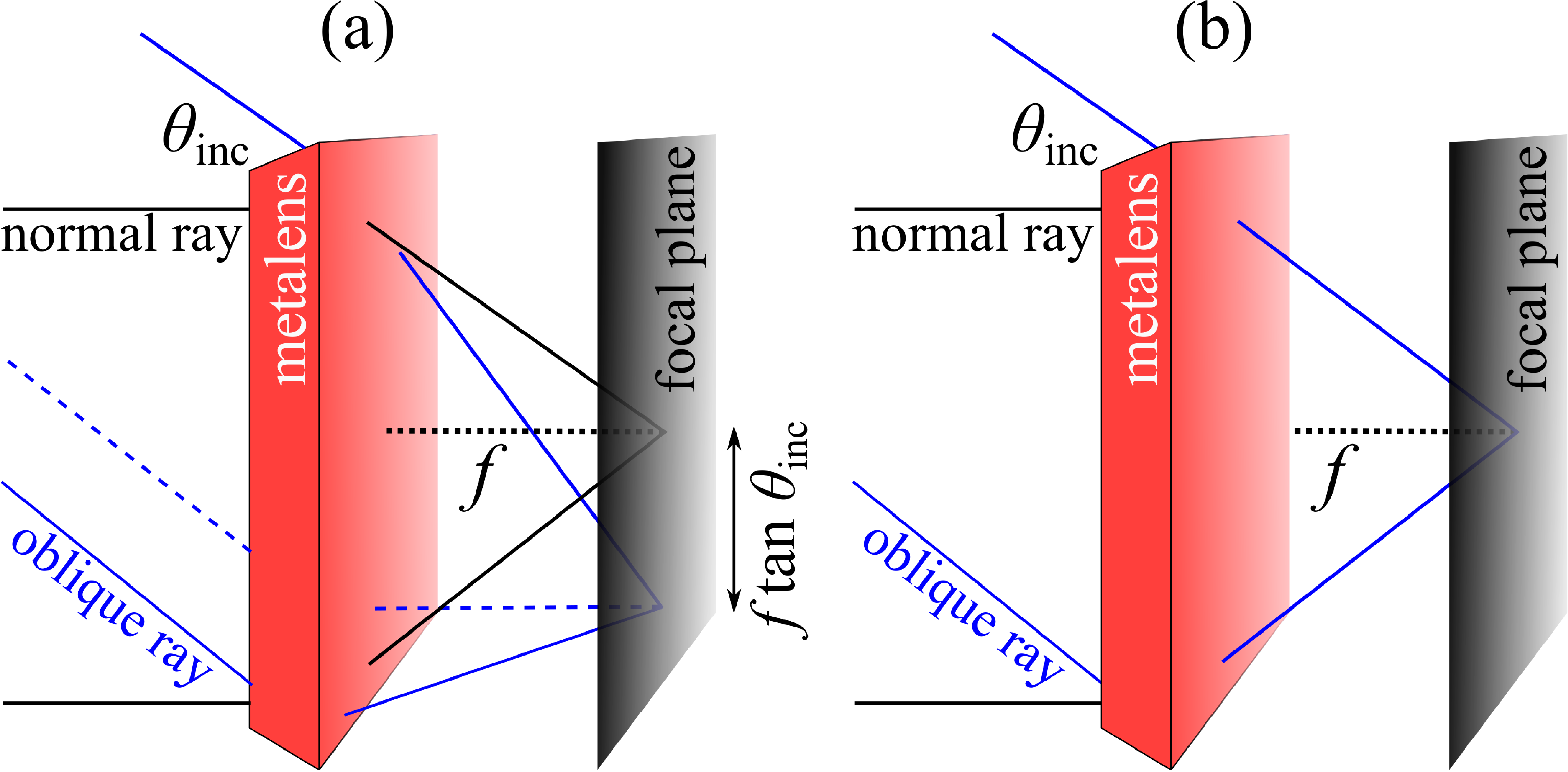}
\caption{Schematics (not to scale) of (a) single-piece nanophotonic
  aberration-corrected metalens and (b) angle-convergent metalens. The
  metalens ensures diffraction-limited focusing under general oblique
  incidence $\theta_\text{inc}$ either onto a laterally shifted focal
  spot (a) or onto the same on-axis focal spot (b). \label{fig:fig1} }
\end{figure}

In this letter, we introduce a different class of MMS involving
several, tightly spaced layers which allow richer physical
interactions within and between layers and thereby offer increased
functionalities. The key property of these MMS is that each layer
cannot be treated independently of the other but must be considered
integrally in the design process. Such a consideration often leads to
a greatly extended design space that cannot be handled by traditional
design methods, which rely on pre-compiled libraries of intuitive
geometrical elements. Below, we propose a general topology
optimization (TO) framework that can automatically discover highly
complex MMS with broad functionalities. As a proof of concept, we
present two TO multi-layered geometries exhibiting angular phase
control: a single-piece nanophotonic metalens with angular aberration
correction~(\figref{fig1}a) and an angle-convergent metalens that
focuses light onto the same focal spot regardless of incident
angle~(\figref{fig1}b).

\begin{figure*}[t!]
\centering
\includegraphics[width=1.7\columnwidth]{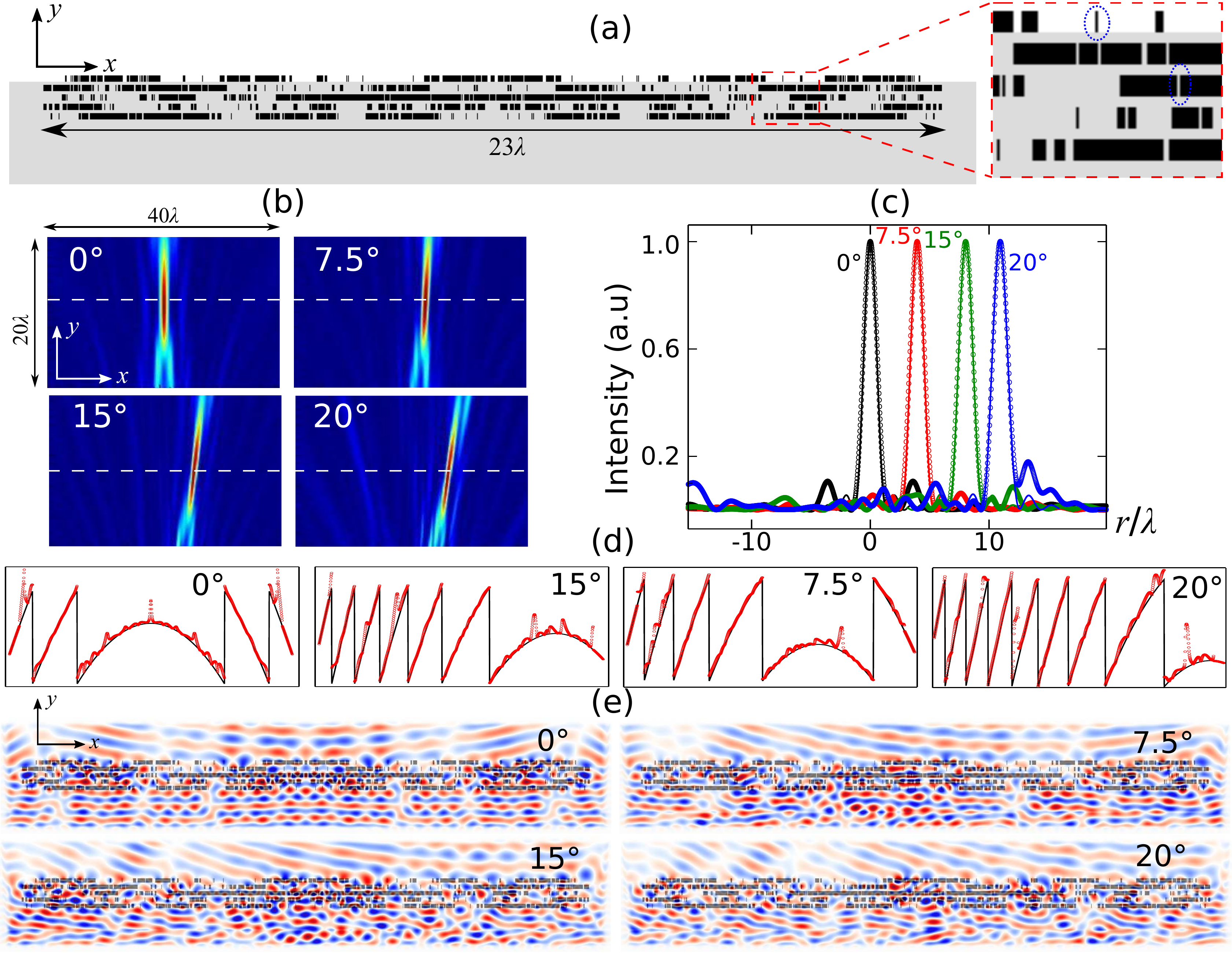}
\caption{(a) Multi-layered miniature 2D lens
  ($\mathrm{NA}=0.35,~f=30\lambda$) which is aberration-corrected for
  four incident angles $\{0^\circ,7.5^\circ,15^\circ,20^\circ\}$. Note
  that by virtue of symmetry, the lens is automatically corrected for
  the negative angles as well
  $\{-7.5^\circ,-15^\circ,-20^\circ\}$. The lens materials consist of
  five layers of silicon (black) in alumina matrix (gray). A portion
  of the lens is magnified for easy visualization (inset); the
  smallest features (such as those encircled within blue dotted oval
  lines) measure $0.02\lambda$ while the thickness of each layer is
  $0.2\lambda$. (b) FDTD analysis of the far field profiles (density
  plots) reveal focusing action for the four incident angles. Note
  that the location of the focal plane is denoted by a white dashed
  line. (c) The field intensities (circle points) at the focal plane
  follow the ideal diffraction limit (solid lines). Note that the
  intensities are normalized to unity for an easy comparison of the
  spot sizes. (d) The corresponding phase profile (red circle data
  points) for each angle is measured at a distance of $1.5\lambda$
  from the device, showing good agreement with the ideal profile
  (black solid line). (e) Near-field profiles with almost perfect
  out-going spherical wavefronts.\label{fig:abfreelens} }
\end{figure*}

 \begin{figure*}[t!]
\centering
\includegraphics[width=1.7\columnwidth]{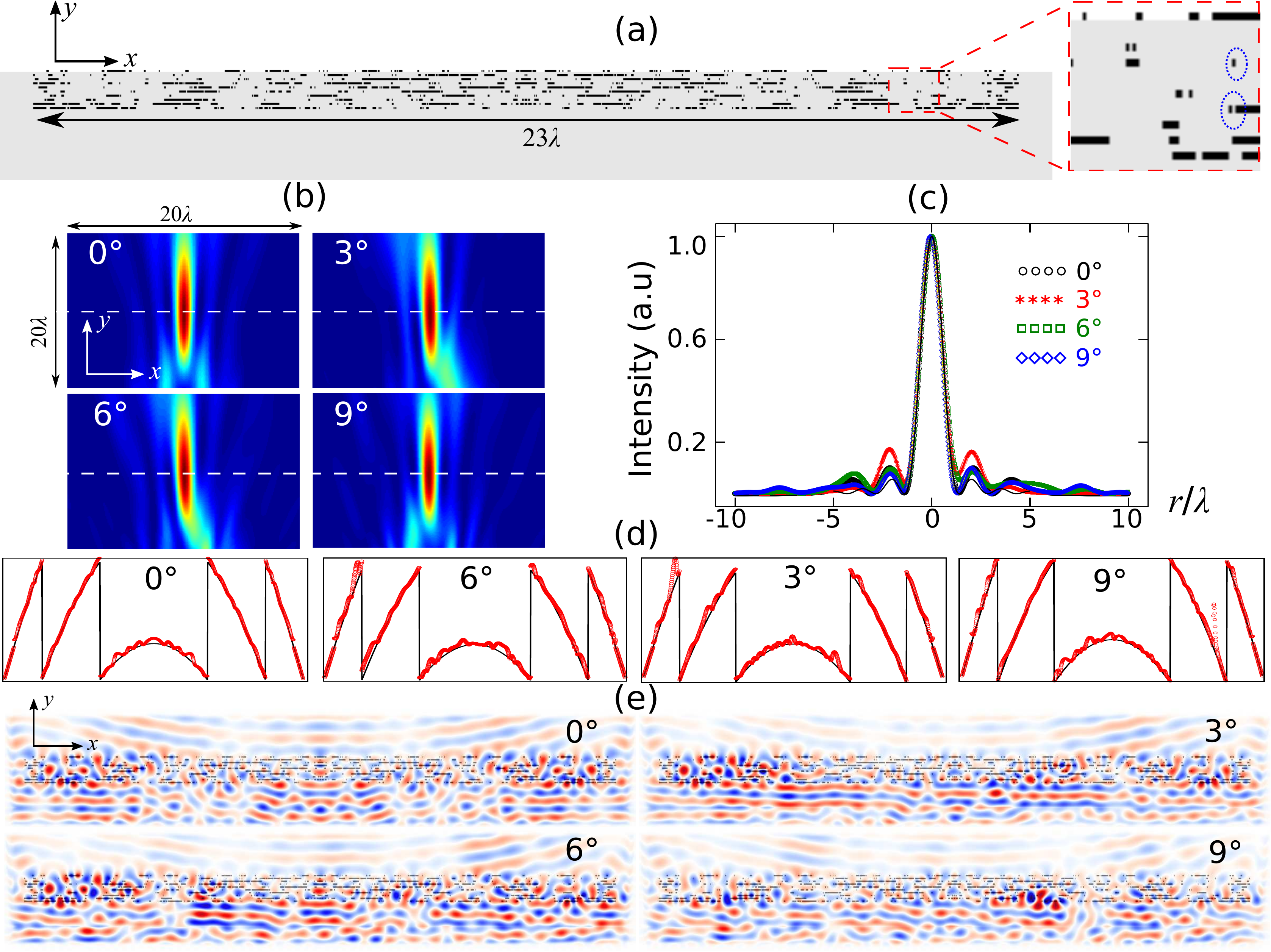}
\caption{(a) Multi-layered miniature 2D lens
  ($\mathrm{NA}=0.35,~f=30\lambda$) that exhibits on-axis focusing for
  the incident angles $\{0^\circ,\pm 3^\circ,\pm 6^\circ,\pm
  9^\circ\}$. The lens materials consist of ten layers of silicon
  (black) in silica matrix (gray). A portion of the lens is magnified
  for easy visualization (inset); the smallest features (such as those
  encircled within blue dotted oval lines) measure $0.02\lambda$ while
  the thickness of each layer is $0.05\lambda$. (b) FDTD analysis of
  the far field profiles (density plots) reveal the same focal spot
  for the different incident angles. Note that the location of the
  focal plane is denoted by a white dashed line. (c) The intensities
  (symbolic data points) at the focal plane follow the on-axis ideal
  diffraction limit for all the incident angles (solid line). (d) The
  corresponding phase profile (red circle data points) for each angle
  is measured at a distance of $1.5\lambda$ from the device, showing
  good agreement with the ideal profile (black solid line). (e)
  Near-field profiles with almost perfect out-going spherical
  wavefronts. \label{fig:sfoclens} }
\end{figure*}

{\it Inverse design formulation.---} Topology optimization (TO) is an
efficient computational technique that can handle an extensive design
space, considering the dielectric permittivity at every spatial point
as a degree of freedom (DOF)~\cite{Yablonovitch05,Jensen11}. A typical
TO electromagnetic problem can be written as:
\begin{align}
\max_{\{\bar{\epsilon}\} } ~ &{\cal F}\left(\mathbf{E};\bar{\epsilon}\right) \\
{\cal G} \left(\mathbf{E};\bar{\epsilon}\right) &\le 0 \\
0 \le \bar{\epsilon} &\le 1.
\end{align}
Here, the DOFs $\{\bar{\epsilon} \}$ are related to the
position-dependent dielectric profile via $\epsilon(\mathbf{r}) =
\left( \epsilon_\text{st} - \epsilon_\text{bg} \right)
\bar{\epsilon}\left(\mathbf{r}\right) + \epsilon_\text{bg}$, where
$\epsilon_\text{st (bg)}$ denotes the relative permittivity of the
structural (background) dielectric material. While $\bar{\epsilon}$
may take intermediate values between 0 and 1, one can ensure a binary
(digital) structure via penalization and filter projection
methods~\cite{Jensen11}. The objective ${\cal F}$ and constraint
${\cal G}$ are often functions of the electric field $\vec{E}$, a
solution of Maxwell's equation,
\begin{align}
  \nabla \times {1 \over \mu}~\nabla \times \vec{E} -~
  \epsilon(\mathbf{r}) {\omega^2 \over c^2} \vec{E} = i \omega \mathbf{J},
\label{eq:ME}
\end{align}
which yields the steady-state $\vec{E}(\mathbf{r};\omega)$ in response
to incident currents $\vec{J}(\mathbf{r},\omega)$ at a given frequency
$\omega$.  While the solution of~\eqref{ME} is straightforward and
commonplace, the key to making optimization problems tractable is to
obtain a fast-converging and computationally efficient adjoint
formulation of the problem. Within the scope of TO, this requires
efficient calculations of the derivatives ${\partial {\cal F} \over
  \partial \bar{\epsilon}(\mathbf{r})},~{\partial {\cal G} \over
  \partial \bar{\epsilon}(\mathbf{r})}$ at every spatial point
$\mathbf{r}$, performed by exploiting the adjoint-variable
method~\cite{Jensen11}.

Recently, inverse-design techniques based on TO have been successfully
applied to a variety of photonic systems including on-chip mode
splitters, nonlinear frequency converters and Dirac cone photonic
crystals~\cite{Jensen11,liang2013formulation,lu2013nanophotonic,men2014robust,piggott2015inverse,shen2015integrated,lin2016cavity,ZinEP3,ZinZIM}. However,
to the best of our knowledge, there is an apparent lack of large-scale
computational techniques specifically tailored for metasurface design,
with the possible exception of Ref.~\cite{JFan17}, which is limited to
grating deflectors. Here, we introduce a general optimization
framework for designing a generic meta-optics device, single or
multi-layered, with arbitrary phase response.  The key to our
formulation is the familiar superposition principle: given a desired
phase profile $\phi(\mathbf{r})$, the ideal wavefront $e^{i
  \phi\left(\mathbf{r}\right)}$ and the complex electric field
$E(\mathbf{r})$ will constructively interfere if and only if their
phase difference vanishes. Defining $E(\mathbf{r}) =
\mathbf{E}(\mathbf{r}) \cdot \mathbf{\hat{e}}$ for a given
polarization $\mathbf{\hat{e}}$, we define the following optimization
function:
\begin{align} {\cal F}(\bar{\epsilon}) = {1 \over V} \int { \Big|
    E(\mathbf{r}) + e^{i \phi(\mathbf{r})} \Big|^2 - \Big|
    E(\mathbf{r}) \Big|^2 - 1 \over 2 \Big| E(\mathbf{r}) \Big| }
  ~d\mathbf{r},
\end{align}
where $V = \int d\mathbf{r}$ and the spatial integration is performed
over a reference plane (typically one or two wavelengths away from the
meta-device) where $\phi(\mathbf{r})$ is defined. Note that ${\cal F}$
is none other than a spatially-averaged cosine of the phase difference
between $e^{i \phi\left(\mathbf{r}\right)}$ and $E(\mathbf{r})$,
\begin{align}
{\cal F}(\bar{\epsilon}) = {1 \over V} \int \cos{\left(\arg{E(\mathbf{r})}-\phi(\mathbf{r})\right)} ~d\mathbf{r},\notag
\end{align}
with the property ${\cal F} \le 1$. Therefore, ${\cal F}$ can be used
to gague and characterize the performance of the device under
construction, with ${\cal F} \approx 1$ indicating that the algorithm
has converged to an optimal solution. In practice, the optimization
algorithms discovered devices with ${\cal F} \approx 99\%$ for many of
the problems under investigation.


{\it Angular phase control.---} An attractive feature of nano-scale
meta-devices is their potential for arbitrary wavefront manipulation
under various control variables including wavelength, polarization or
incident angle. Although spectral and polarization control have been
explored in a number of previous
works~\cite{aieta2015multiwavelength,mueller2017metasurface}, to the
best of our knowledge, angular control has not been achieved so
far. In fact, realizing angular control in traditional single-layer
ultra-thin metasurfaces might prove fundamentally impossible since the
interface is constrained by the generalized Snell's
laws~\cite{yu2011light}. On the other hand, MMS with thicknesses on
the order of a wavelength or more (whose internal operation cannot be
described via ray optics) can overcome such a limitation; in
principle, they can be engineered to exhibit directionality even
though conventional approaches which rely on intuitive, hand designs
might prove unequal to such a task. Here, we leverage our optimization
algorithm to develop multi-functional structures where an arbitrary
phase response that varies with the angle of incidence can be
imprinted on the same device.

The traditional objective in the design of metalenses is the creation
of a single, hyperbolic phase profile, $\phi(r)= \phi_0 - {2 \pi \over
  \lambda} \left(\sqrt{f^2 + \left(r - r_0\right)^2} - f\right)$,
characterized by the focal length $f$, in response to a normally
incident plane wave~\cite{aieta2015multiwavelength}. Here, $r_0$
denotes the center of the lens whereas $\phi_0$ denotes an arbitrary
phase reference that can be varied as an additional degree of freedom
in the metasurface design~\cite{khorasaninejad2017achromatic}. As
discussed in the Ref.~\cite{aieta2013aberrations}, such a design is
free of spherical aberrations but still suffer from angular and
off-axis aberrations such as coma and field curvature. These errors
arise out of an incorrect phase profile that skews the oblique
off-axis rays. A corrected phase profile free from aberration is
therefore necessarily angle-dependent, as given by:
\begin{equation}
\resizebox{1\columnwidth}{!}{$\phi\left(r,\theta_\text{inc}\right)=\phi_0(\theta_\text{inc})-{2 \pi \over \lambda} \left(\sqrt{f^2 + \left(r - r_0 - f \tan{\theta_\text{inc}}\right)^2} - f\right).$} \notag
\end{equation}
Note that the above expression can be deduced by considering the
optical path length contrast between a generic ray and the orthonormal
ray directed towards a focusing spot laterally shifted by $f
\tan{\theta_\text{inc}}$ (see \figref{fig1}a, blue dashed line). Here,
we leverage our TO algorithm to design a 2D miniature
\emph{angle-corrected metalens} that exactly embodies the ideal
angle-dependent phase profile given above. Note that though our
miniature design is a proof-of-concept theoretical prototype, it is
completely straightforward (though computationally intensive) to
design a full 3D wide-area (centimeter-scale) single-piece
monochromatic aberration-free lens using our TO technique. We
emphasize that such a ``next generation" lens fundamentally differs
from the traditional aberration-corrected doublet because the latter
exclusively relies on classical ray tracing techniques whereas the
former intricately exploits nano-scale electromagnetic effects to
achieve angular control.

We design a lens with an NA of 0.35 and a focal length of
$30\lambda$. The device consists of five layers of topology-optimized
aperiodic silicon gratings (invariant along $z$) against amorphous
alumina background~(\figref{abfreelens}a). Each silicon layer is
$0.2\lambda$ thick and is separated by $0.1\lambda$ alumina gaps. We
specifically chose silicon and alumina with a view to eventual
fabrication at mid or far IR wavelengths ($5 - 8~\mathrm{\mu m}$) by
stacking patterned 2D slabs via repeated lithography, material
deposition and planarization processes~\cite{3d2,3d4}. The entire lens
has a thickness of $1.5\lambda$, offering ample space for complex
electromagnetic interactions while, at the same time, maintaining
orders of magnitude smaller thickness compared to traditional
multi-lens systems. The lens is aberration corrected for four incident
angles $\{0^\circ,7.5^\circ,15^\circ,20^\circ \}$ as well as their
negative counterparts $\{-7.5^\circ,-15^\circ,-20^\circ \}$. Note that
the largest possible angle for diffraction-limited focusing is
$\approx 21^\circ$ and is determined by the numerical aperture. For
simplicity, we consider off-axis propagation in the $xy$ plane with an
$s$-polarized electric field parallel to the direction of the
gratings, $\mathbf{E}=E(\mathbf{r}) \hat{z}$. FDTD analysis of the far
field~(\figref{abfreelens}b) reveals focusing action with diffraction
limited intensity profiles~(\figref{abfreelens}c), while the
transmission efficiencies average around $25\%$ for the four
angles. To evaluate the deviation of our design from the ideal phase
profile, we computed the wave aberration function (WAF) for each
angle~\cite{aieta2013aberrations}, obtaining $\mathrm{WAF}(0^\circ,\pm
7.5^\circ,\pm 15^\circ,\pm 20^\circ)= (0.07,0.04,0.06,0.08)$, which
clearly satisfy the Mar\`{e}chal criterion $\mathrm{WAF} \le {1 \over
  14}$ except for the $20^\circ$ incident angle. The errors in the
latter case primarily arise from the difficulty over optimizing the
extremities of the lens, which can be mitigated by extending the
optimized lens area (or equivalently designing a larger NA). It is
worth noting that the residual phase errors apparent in the optimized
design primarily stem from a need to force the optimal design to be
binary while being constrained by a limited resolution. In this work,
we implemented a spatial resolution step size $\Delta r = \lambda/50$
over a $23\lambda$-long simulation domain while our optimization
algorithm handles approximately 5600 degrees of freedom. These
parameters are solely dictated by the limited computational resources
currently available to us. We find that without the binary constraint
(i.e. when each DOF is allowed to take intermediate values between 0
and 1), the optimal designs easily achieve perfect phase profiles with
WAFs smaller than $0.01$. We expect that given better computational
facilities, optimization over higher resolution domains will lead to
fully binary structures that also preserve vanishing
$\mathrm{WAF}\approx 0$.
 
Next, to demonstrate the versatility of our approach, we design a 2D
metalens that focuses light onto the same spot regardless of the angle
of incidence (\figref{fig1}b) -- a device which we will choose to call
\emph{angle-convergent metalens}. Specifically, we impose the phase
profile $\phi(r)=\phi_0\left(\theta_\text{inc}\right) - {2 \pi \over
  \lambda} \left(\sqrt{f^2 + \left(r - r_0\right)^2} - f\right)$ on
the outgoing field under multiple discrete incident angles
$\{0^\circ,\pm 3^\circ,\pm 6^\circ,\pm 9^\circ\}$. The lens has an NA
of 0.35 and a focal length of $30\lambda$.  The lens materials consist
of ten layers of $0.05\lambda$ thick silicon in silica separated by
$0.05\lambda$ gaps~(\figref{sfoclens}a), making the entire device
approximately one $\lambda$ thick. Such a device can be fabricated
using advanced 3D photonic integration techniques~\cite{3d2},
including those enabled by CMOS foundries~\cite{3d1}. Far field
analysis~(\figref{sfoclens}b) shows focusing action at the same focal
spot for all the angles. Although the field intensities at the focal
spot do not exactly follow the profile of an ideal Airy disk due to
residual phase errors, their bandwidth (aka full width at half
maximum) clearly satisfies the diffraction
limit~(\figref{sfoclens}c). The diffraction-limited focusing is also
consistent with small WAFs which are found to satisfy the Mar\`{e}chal
criterion: $\mathrm{WAF}(0^\circ,\pm 3^\circ,\pm 6^\circ,\pm 9^\circ)=
(0.02,0.04,0.04,0.02) < 1/14$. The transmission efficiency of the
device averages around 15\% over all angles.
 
{\it Conclusion and outlook.---} To summarize, we proposed a general
optimization framework for inverse design of multi-layered
meta-optics. We leveraged our formulation to engineer angular phase
control in multi-layered metalens. It is important to note that, in
this paper, as we focus on establishing the validity and versatility
of our optimization approach, we have not sought to pursue ``the best
possible design'' for any particular problem that we chose to
investigate. For example, the number, positioning and thicknesses of
layers are arbitrarily chosen in each problem. It is entirely possible
that depending on the desired level of performance, one can achieve
viable designs using fewer and/or thicker layers, which could render
the entire device even thinner and easier to fabricate.

While the optimization framework we have proposed exclusively focuses
on phase, work is currently under way to implement additional features
such as amplitude uniformity and high efficiency constraints, which
can be straightforwardly added to our formulation. Although the
addition of extra conditions would presumably strain the optimization
process, we expect that a full 3D multi-layered device platform should
be able to accommodate any additional demands. Ultimately, we surmise
that multi-layered volumetric structures (no more than a few
wavelengths thick) will help deliver unprecedented wavefront
manipulation capabilities at the nano-scale that involve phase,
intensity and polarization control as well as spectral and angular
dispersion engineering \emph{altogether} in a single device. The TO
technique is by far the most efficient tool that can handle the
enormous design space available to such platforms. Although
fabrication of multi-layered nanostructures might prove challenging
for shorter operational wavelengths, they can be readily implemented
in mid to far IR regimes, through state-of-the-art 3D fabrication
technologies~\cite{3d2} such as two-photon lithography~\cite{3d5} and
laser writing processes~\cite{3d3}, advanced foundry access~\cite{3d1}
as well as ultra-high resolution EUV lithography~\cite{euv}.

{\it Acknowledgements.---} We would like to thank Raphael Pestourie,
Steven Johnson, Pawel Latawiec, Robert Devlin, Wei-Ting Chen and
Alexander Zhu for useful discussions. Z. Lin is supported by the
National Science Foundation Graduate Research Fellowship Program under
Grant No. DGE1144152. The work was partially supported by the Air
Force Office of Scientific Research (MURI grant no. FA9550-14-1-0389),
by the National Science Foundation under Grant No. DMR-1454836, by the
National Science Foundation under EFRI-1005093 and by the Princeton
Center for Complex Materials, a MRSEC supported by NSF Grant No. DMR
1420541.

\bibliographystyle{unsrt}
\bibliography{refs}

\beginsupplement

\section{Supplement}

\section{Phase profile of an ideal aberration-free lens}
Here, we examine the angle-dependent phase profile necessary to correct monochromatic aberrations in a lens:
\begin{align}
\phi\left(r,\theta_\text{inc}\right)=c(\theta_\text{inc})-{2 \pi \over \lambda} \left(\sqrt{f^2 + \left(f \tan{\theta_\text{inc}} - r + r_0\right)^2} - f\right).
\end{align}
Here, $r_0$ is the origin of the lens. It is easily seen that the aberration-corrected $\phi\left(r,\theta_\text{inc}\right)$ is simply a hyperbolic profile shifted by a distance $f \tan{\theta_\text{inc}}$ which is the location of focal spot in the far field. When the angle of incidence $\theta_\text{inc}$ is zero, we obtain the familiar hyperbolic phase profile for the normal incidence. Given $\phi(r)$, we can exactly compute the far field by convoluting with the standard Green's function propagator:
\begin{align}
E_\text{far}\left(\mathbf{r},\theta_\text{inc}\right) \propto \int { e^{i k |\mathbf{r}-\mathbf{r}'|} \over |\mathbf{r}-\mathbf{r}'| } e^{i \phi\left(r',\theta_\text{inc}\right)} d\mathbf{r}' \label{eq:eqff}
\end{align}
In our paper, we have exactly computed the ideal far field intensity profile based on the convolution integral above (Eq.~\ref{eq:eqff}) and compared it with the simulated far field of the optimized design for each incident angle. It should be noted that for $\sin{\theta_\text{inc} } \gtrsim \mathrm{NA}$, the width (FWHM) of the far field begins to violate the diffraction limit $> {\lambda \over 2 \mathrm{NA}}$.  

\section{Comparison against standard normal-incidence metalens design}
\begin{figure*}[b!]
\centering
\includegraphics[width=\textwidth]{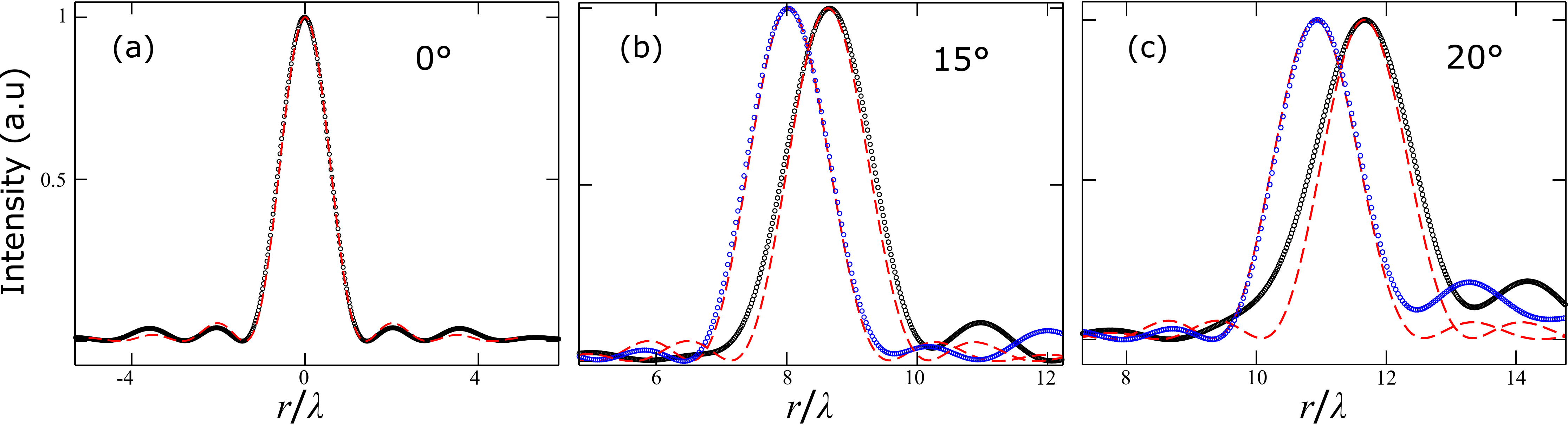}
\caption{Comparison of performance between angle-corrected metalens and the standard normal-incidence metalens. We examine a conventional single-layered metalens ($\mathrm{NA}=0.35,~f=30\lambda$) optimized for diffraction-limited focusing at normal incidence $0^\circ$. (a) Black circle data points denote the far field intensity at the focal plane of the standard lens whereas the red dashed line represents the ideal intensity profile. (b) The far field intensity profile (black circle data points) for $15^\circ$ exhibits discernible deviations from the ideal diffraction limit (red dashed lines): shifted field maximum, line-width broadening, skewed profile. Note that a displaced replica of the ideal profile is superimposed on the black circle data points for easy comparison of the linewidths. Blue circle data points represent the field profile of the topology-optimized angle-corrected meta-lens. (c) The deviation of the standard lens from the ideal scenario is even more pronounced for the larger incident angle of $20^\circ$. \label{fig:comp} } 
\end{figure*}

 We examine the standard single-layered metalens ($\mathrm{NA}=0.35,~f=30\lambda$) with the same parameters as our topology-optimized aberration-corrected design given in the main text. Precisely speaking, a standard lens is one which is optimized only for diffraction-limited focusing at the normal incidence $0^\circ$. \Figref{comp}a (black circle data points) shows that the standard lens ensures perfect diffraction limited focusing along the ideal Airy profile (red dashed line). For $15^\circ$ incident angle (\figref{comp}b), the field maximum (black circle data points) of the standard lens deviates from the ideal spot on the focal plane while exhibiting slight broadening and marked asymmetry. Note that a displaced replica of the ideal profile (red dashed line) is superimposed on the black circle data points for an easy comparison of the two linewidths between the standard and ideal lenses. Blue circle data points represent the field profile of the topology-optimized angle-corrected meta-lens which almost perfectly follows the ideal profile. Deviations in the standard lens are even more pronounced for the larger incident angle of $20^\circ$ (\figref{comp}c).

\section{Angle Sweep}
\begin{figure*}[h!]
\centering
\begin{subfigure}{.5\textwidth}
  \centering
  \includegraphics[width=.9\linewidth]{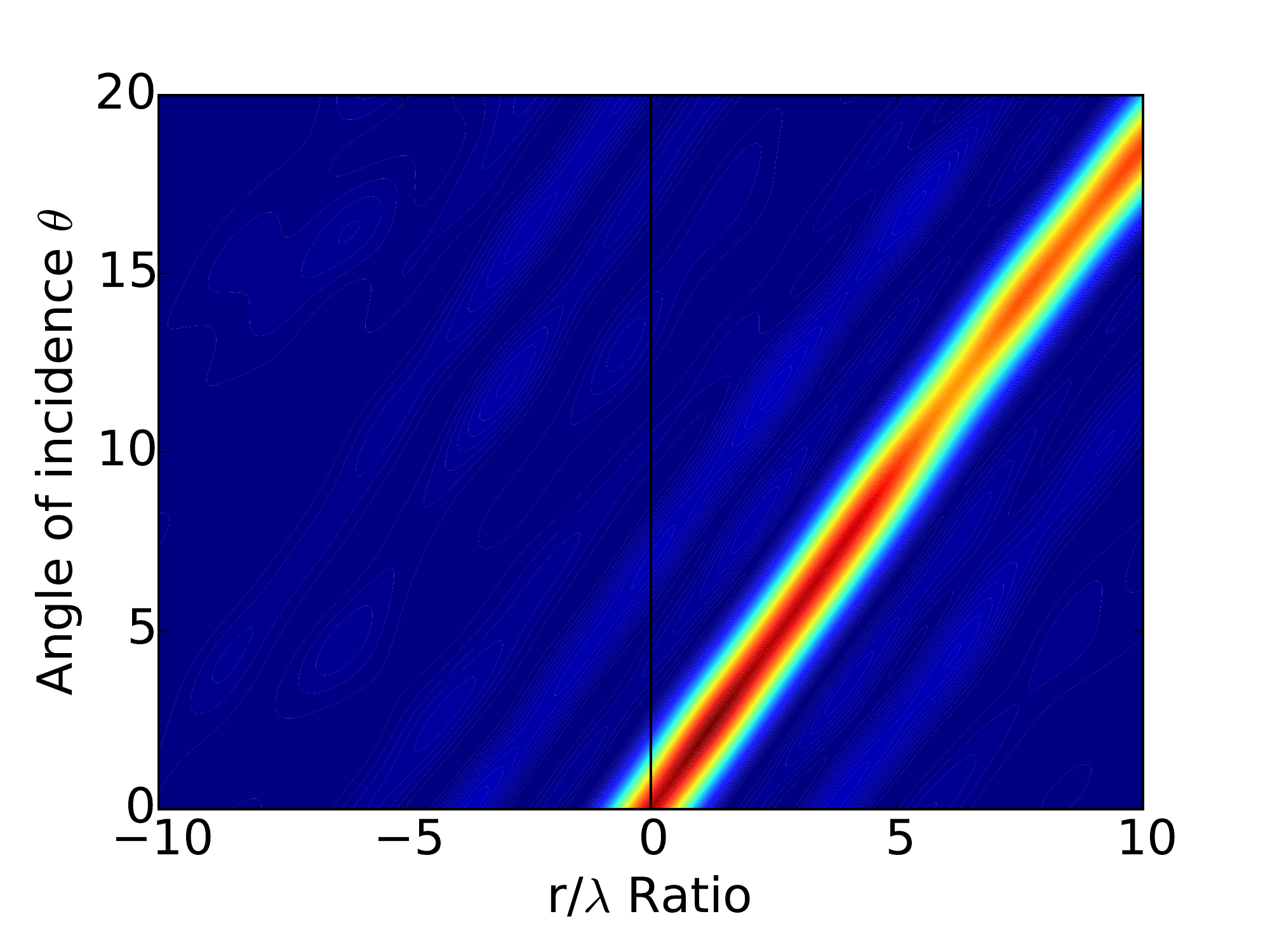}
  \caption{Aberration-corrected meta-lens}
  \label{fig:AngleSweepAbberationML}
\end{subfigure}%
\begin{subfigure}{.5\textwidth}
  \centering
  \includegraphics[width=.9\linewidth]{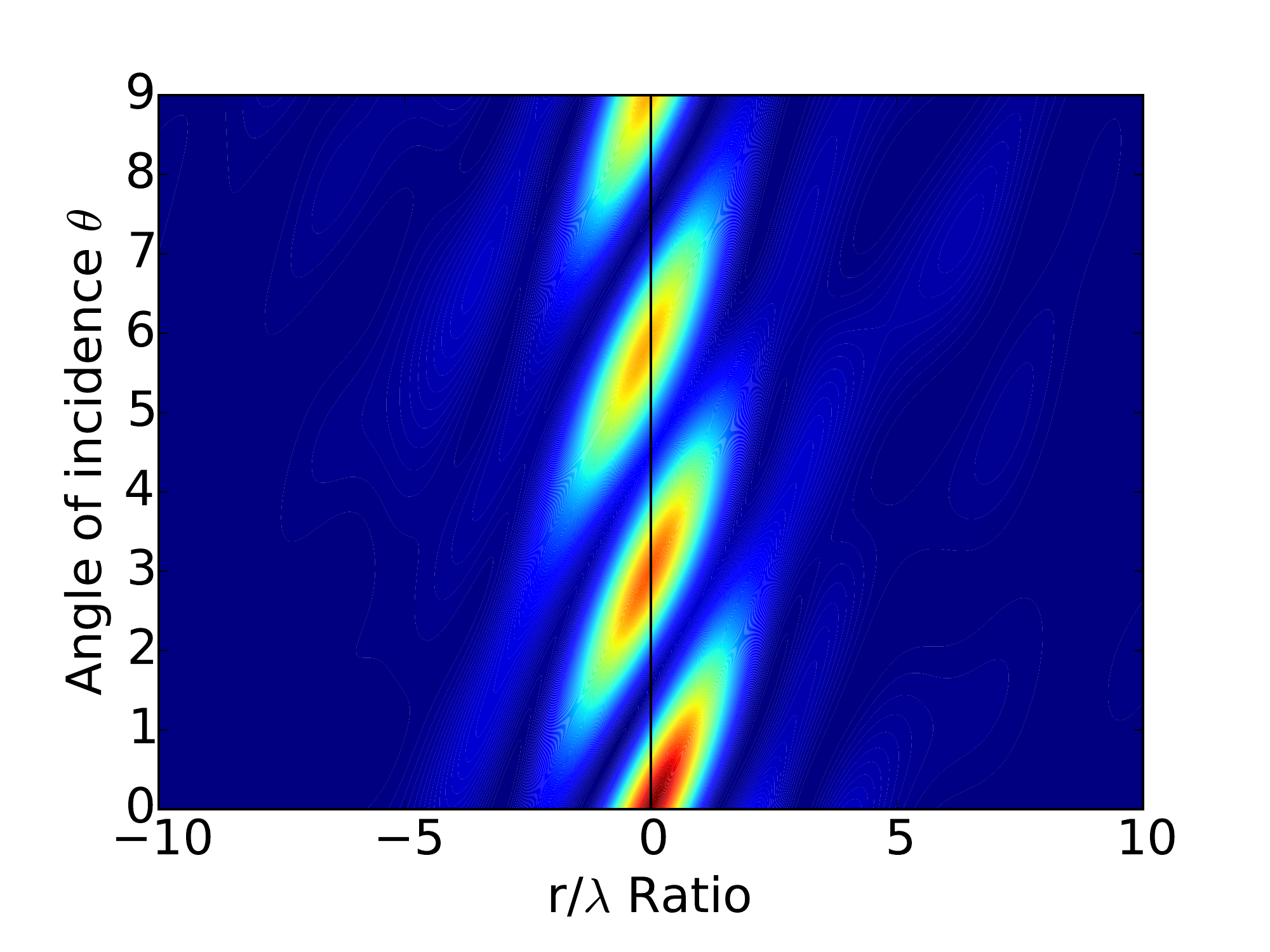}
  \caption{On axis focusing}
  \label{fig:AngleSweepOnAxisFocusing}
\end{subfigure}
\caption{Focal spot intensities for (a) the aberration-corrected meta-lens and the (b) the on axis focusing lens for different angles of incidence}
\label{fig:anglesweep}
\end{figure*}

In \Figref{anglesweep}, we examine the topology-optimized designs for different angles of incidence. The aberration-corrected meta-lens (\Figref{AngleSweepAbberationML}) has a continuous focusing function while the angle-convergent on-axis focusing meta-lens has discrete focal spots at the optimized angles of incidence (\Figref{AngleSweepOnAxisFocusing}). We expect the result can be further improved by optimizing many more angles.

\section{Detailed structures of the topology-optimized meta-lenses and field profiles}
\begin{figure*}[t]
\makebox[\textwidth][c]{\includegraphics[width=1.15\textwidth]{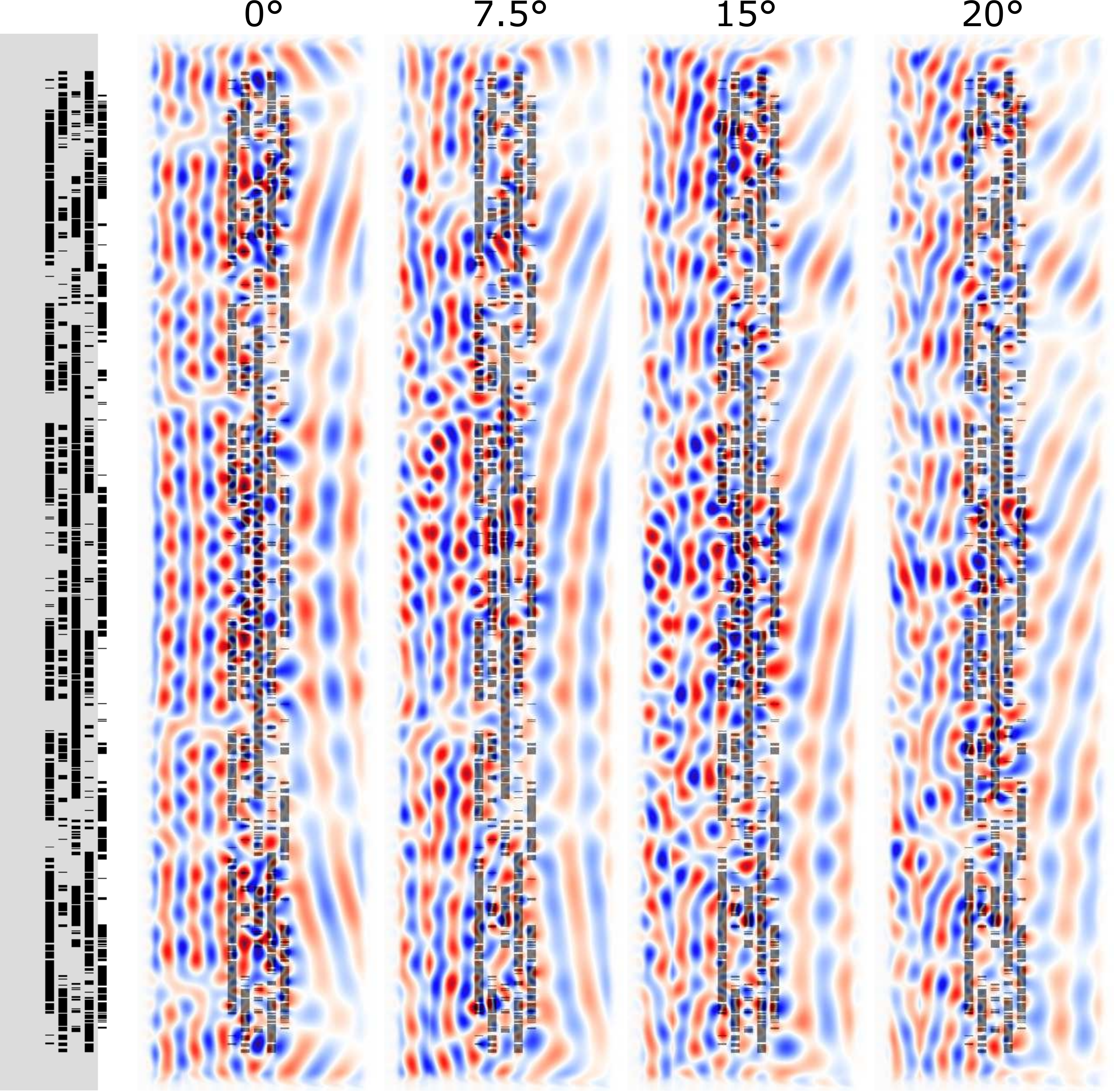}}
\caption{Angle-corrected metalens and near-field profiles\label{fig:abelens}}
\end{figure*}

\begin{figure*}[t]
\makebox[\textwidth][c]{\includegraphics[width=1.15\textwidth]{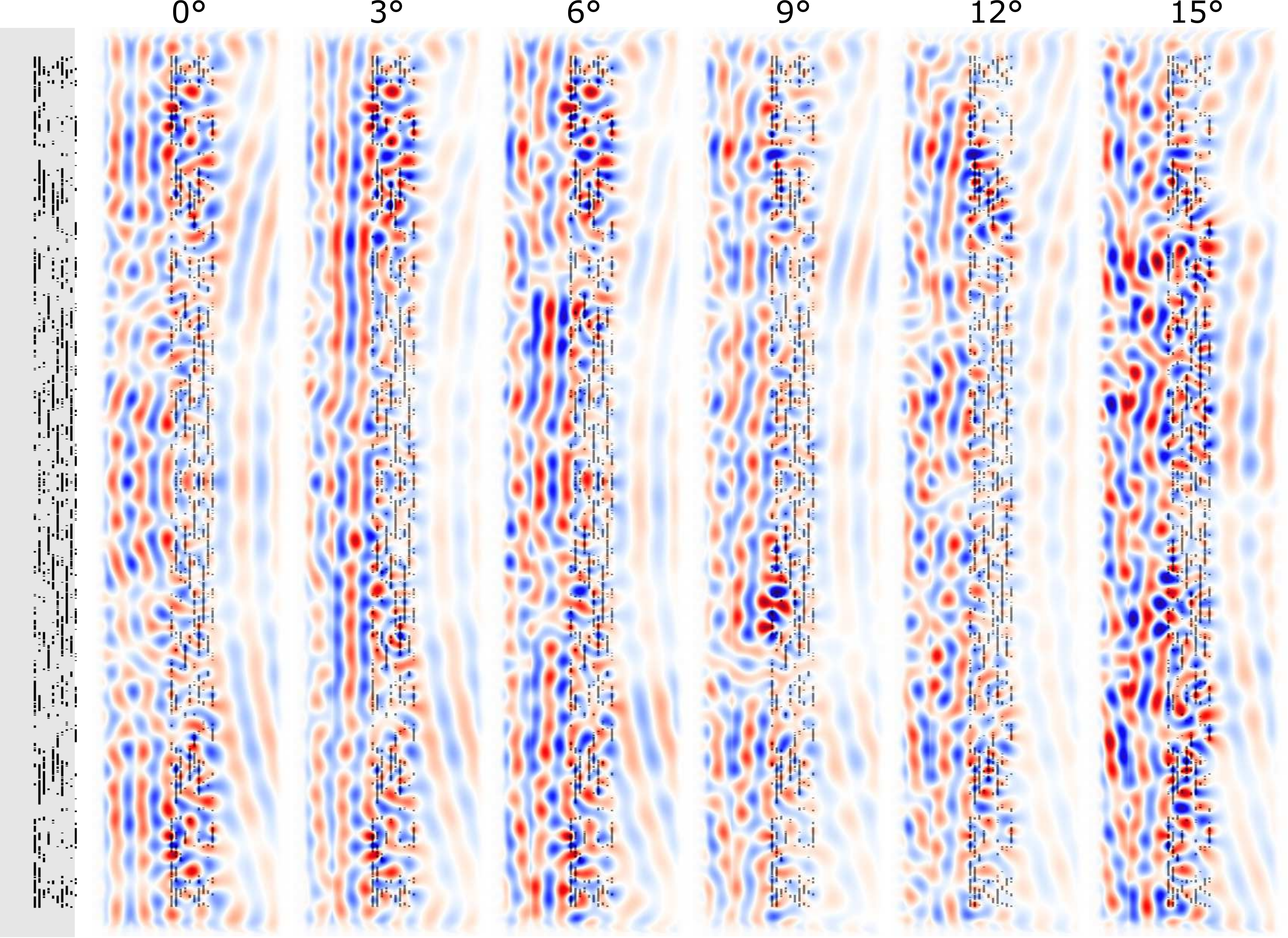}}
\caption{Angle-convergent metalens and near-field profiles\label{fig:convlens}} 
\end{figure*}

Detailed structural images of the angle-corrected and angle-convergent metalenses are given in \figreftwo{abelens}{convlens}, together with the corresponding ``out-of-page" $s$-polarization near-field profiles showing almost perfect spherical wavefronts. Note that the design of the angle-convergent metalens~(\figref{convlens}) has been tweaked to accommodate on-axis focusing at the larger angles of $12^\circ$ and $15^\circ$. Although we achieve diffraction-limited focusing for these angles as well (not shown), the transmission efficiencies become much smaller $\sim 3\%$. In future works, this issue will be fixed by explicitly incorporating a high-transmission efficiency constraint into the optimization process.

\end{document}